\documentclass[aps,twocolumn,showpacs]{revtex4}
\usepackage{bbm}
\usepackage{amssymb}
\usepackage{amsfonts}
\usepackage{amsmath}
\usepackage{mathrsfs}

\begin{document}

\title{Art of spin decomposition}

\author{Xiang-Song Chen$^{1,2}$}
\email{cxs@hust.edu.cn}
\author{Wei-Min Sun$^{3,2}$}
\author{Fan Wang$^{3,2}$}
\author{T. Goldman$^4$}

\affiliation{$^1$Department of Physics, Huazhong
University of Science and Technology, Wuhan 430074, China\\
$^2$Kavli Institute for Theoretical Physics China, Chinese
Academy of Science, Beijing
100190, China\\
$^3$Department of Physics, Nanjing University, and Joint
Center for Particle, Nuclear Physics and Cosmology, Nanjing
210093, China\\
$^4$Theoretical Division, Los Alamos National Laboratory,
Los Alamos, NM 87545, USA}

\date{\today}

\begin{abstract}

We analyze the problem of spin decomposition for an interacting
system from a natural perspective of constructing angular momentum
eigenstates. We split, from the total angular momentum operator, a
proper part which can be separately conserved for a stationary
state. This part commutes with the total Hamiltonian and
thus specifies the quantum angular momentum. We first show
how this can be done in a gauge-dependent way, by seeking a specific
gauge in which part of the total angular momentum operator vanishes
identically. We then construct a gauge-invariant operator with the
desired property. Our analysis clarifies what is the most pertinent
choice among the various proposals for decomposing the nucleon spin.
A similar analysis is performed for extracting a proper part from the
total Hamiltonian to construct energy eigenstates.

\end{abstract}
\maketitle

How the nucleon spin originates from its internal
quark-gluon dynamics appears to be a rather complicated and
challenging problem. \cite{SpinReview} A reflection of the
complication is that even a schematic decomposition of the
nucleon spin is not agreed upon. So far,  quite a few
proposals for decomposing the nucleon spin have appeared,
from the early popular schemes of Jaffe-Manohar
\cite{Jaff90} and Ji \cite{Ji97}, to the recent
gauge-invariant schemes of Chen {\it et al.}
\cite{Chen08,Chen09,Chen11}, Wakamatsu \cite{Waka10}, Cho
{\it et al.} \cite{Cho10}, and Leader \cite{Lead11} (to
list just a few). This frequently causes great confusion,
when one piece (say, gluon spin) from a certain scheme is
examined together with another piece (say, quark orbital
angular momentum) from another scheme, and makes people
feel lost as to which decomposition is better.

To clarify the issue, we point out a seldom noticed fact that
decomposing the total angular momentum of an interacting system is
actually a common practice in quantum mechanics. The same occurs
also for the Hamiltonian. These decompositions, however, indeed
involve a rather delicate art, which apparently has never been
properly addressed. Revealing this art provides a clear clue
as to how to decompose the nucleon spin most naturally.

To see the point, let us look at the quantum-mechanical study of the
Hydrogen atom (an interacting system of electron,
proton, and electromagnetic field). The standard procedure is to
construct Hamiltonian and angular momentum eigenstates of {\em only}
the electron:
\begin{eqnarray}
i\partial_t \psi_e =H_e \psi_e =E_e \psi_e\\
\vec J_e^2 \psi_e =j(j+1) \psi_e,~ ~ J^z_e \psi_e =m \psi_e,
\end{eqnarray}
where
\begin{eqnarray}
H_e=\vec \alpha \cdot \frac 1i \vec
D_e+M_e\beta +q_eA^0 \\
\vec J_e=\frac 12 \vec \Sigma + \vec x\times
\frac 1i \vec
\partial
\end{eqnarray}
are the electron Hamiltonian and angular momentum operators,
respectively. $\vec D_e =\vec \partial -iq_e \vec A$ is the
covariant derivative for the electron, with $q_e$ the electron
charge. The {\em electron} quantum numbers $E_e$, $j$ and $m$ are
then used to label an {\em atomic} state. But one may seriously ask:
how can $E_e$, $j$ and $m$ represent the energy and angular momentum
of the atom, when only the following {\em total atomic} Hamiltonian
and angular momentum are conserved?
\begin{eqnarray}
H_{\rm atom} &=& \int d^3 x \psi_e^\dagger (\vec \alpha \cdot \frac
1i
\vec D_e+M_e\beta) \psi_e  \nonumber \\
&&+\int d^3 x \psi_p^\dagger (\vec \alpha \cdot \frac 1i \vec
D_p+M_p\beta ) \psi_p \nonumber \\
&&+\int d^3 x \frac 12 (\vec E^2+\vec B^2) \label{Ha} \\
\vec J_{\rm
atom} &=& \int d^3 x \psi_e ^\dagger (\frac 12 \vec
\Sigma +\vec x\times  \frac 1i \vec \partial)\psi_e \nonumber \\
&&+\int d^3 x \psi_p ^\dagger (\frac 12 \vec \Sigma +\vec x\times
\frac 1i \vec \partial)\psi_p \nonumber\\
&&+\int d^3x (\vec E \times \vec A+
E_i \vec x\times \vec \partial A_i)\nonumber \\
&\equiv& \vec J_e+\vec J_p+\vec J_\gamma \label{Ja}
\end{eqnarray}
(We use the same symbols for quantum-mechanical and quantum-field
operators, since confusion can hardly arise. $\psi_e$ is the
electron field, $\psi_p$ is the proton field.)

A careful reader would notice that $H_e$ and $\vec J_e$ are
even gauge dependent and so have no definite contents at
all! For these operators to be useful, a key role must be
played by the choice of gauge. Indeed, we show by a careful
examination that, under certain circumstance and in a
special gauge, the use of electron quantum numbers $E_e$,
$j$ and $m$ for the total atom can be justified.

The circumstance we consider is a {\em stationary} system,
namely, the electric current $j^\mu=q_e\bar \psi_e \gamma
^\mu \psi_e+q_p\bar \psi_p \gamma ^\mu \psi_p$ and the
electromagnetic fields $F^{\mu\nu}$ are {\em
time-independent.} This includes the typical case of
solving for the quantum-mechanical eigenfunctions. Both
$j^\mu$ and $F^{\mu\nu}$ are gauge-invariant quantities, so
their time-independence has a definite physical meaning. In
contrast, the electron wavefunction $\psi_e$ and the
electromagnetic vector potential $A^\mu$ are
gauge-dependent. Such a gauge-dependence can be both a
disadvantage and an advantage, which are just two sides of
the same coin. The disadvantage, as we remarked above, is
that the individual angular momentum operators in Eq.
(\ref{Ja}), $\vec J_e$, $\vec J_p$, and $\vec J_\gamma$ are
all gauge-dependent; therefore an electron angular momentum
eigenstate may have no definite physical meaning. (A more
serious and tricky problem is the gauge dependence of
$H_e$, which we address shortly below.) The advantage, on
the other hand, is that there must always exist a gauge in
which one element of the electromagnetic angular momentum
(say, $ J^z_\gamma$) vanishes, therefore $J^z _e+J^z_p$ in
this gauge equals the total $J^z_{\rm atom}$ and thus can
be in an eigenstate. A non-trivial and remarkable feature,
however, is that for a stationary system, a single gauge
condition can lead to the vanishing of all three components
of $\vec J_\gamma$.

{\em Proof:} As $\partial_t F^{\mu\nu}=0$, we have
$\vec \partial \times \vec E =-\partial_t \vec B=0$, hence $\vec E$
must be a gradient, which we denote as $-\vec \partial \phi$. Then
a little algebra shows that
\begin{eqnarray}
\vec J_\gamma ^{\rm stat}&=& \int d^3x (-\vec\partial \phi) \times
\vec A+ \int d^3x \vec x\times
(-\partial_i \phi) \vec \partial A_i \nonumber \\
&=&\int d^3x \phi \vec x \times \vec \partial (\partial_i A_i).
\end{eqnarray}
Thus, $\vec J_\gamma ^{\rm st}\equiv \vec 0$ in the Coulomb gauge
$\partial _i A_i=0$. In this gauge, therefore, not only
$J^z_p+J^z_e$, but also $(J^z_p+J^z_e)^2$, can be in an eigenstate.
Furthermore, if the proton (with a magnetic moment much smaller
than that of the electron) is unpolarized, then the electron quantum
numbers $j$ and $m$ do represent the angular momentum of the total
atom.

The use of Coulomb gauge appears to be taken-for-granted in quantum
mechanics. From our analysis, it is truly a very fortunate choice:
Should other gauges be chosen, one would have to account for a
(spuriously) non-zero electromagnetic angular momentum so as to
obtain the correct total atomic spin. (Unfortunately, such good fortune
is not always cherished: We will see soon that the study of nucleon
spin structure involves  a great deal of effort in exploring {\em spurious}
gluon angular momentum.)

One can further appreciate the good fortune in choosing
Coulomb gauge by considering the Hamiltonian. In Eq.
(\ref{Ha}), we have written the total Hamiltonian in a most
familiar, explicitly gauge-invariant form, where $\frac 12
(\vec E^2+\vec B^2)$ is the usual expression for energy of
the electromagnetic field in classical electrodynamics. But
the form in Eq. (\ref{Ha}) is neither useful nor
illuminating in quantum mechanics. First of all, $H_e$ does
not show up explicitly in Eq. (\ref{Ha}). Moreover, it is
not clear from Eq. (\ref{Ha}) whether the parts other than
$H_e$ (i.e., $H_{\rm atom}-H_e$) can be ignored for an
atom. For the sake of justifying the ``electron energy'',
$E_e$, as the pertinent label of an atomic energy level, it
is better to put the total atomic Hamiltonian in the
canonical form (with gauge-variant densities):
\begin{eqnarray}
H_{\rm atom} &=&\int d^3x(\psi_e^\dagger i\partial_t \psi_e
+\psi_p^\dagger i\partial_t \psi_p -E^i
\partial_t A^i -{\mathscr L})\nonumber \\
&=&\int d^3 x \psi_e^\dagger (\vec \alpha \cdot \frac 1i \vec
D_e+M_e\beta +q_e A^0) \psi_e \nonumber\\
&+&\int d^3 x \psi_p^\dagger (\vec \alpha \cdot \frac 1i
\vec
D_p+M_p\beta +q_pA^0) \psi_p\nonumber \\
&-&\int d^3 x [E^i\partial_t A^i +\frac 12 (\vec E^2-\vec B^2)],
\end{eqnarray}
where the Lagrangian is
\begin{equation}
{\mathscr L}=\bar \psi_e (i\gamma_\mu D^\mu_e-M_e)\psi_e +\bar
\psi_p (i\gamma_\mu D^\mu_p-M_e)\psi_p -\frac 14 F^{\mu\nu}
F_{\mu\nu}
\end{equation}
By some careful algebra, we obtain
\begin{eqnarray}
H_{\rm atom} = \int d^3 x \psi_e^\dagger (\vec \alpha \cdot \frac 1i
\vec D_e+M_e\beta +q_eA^0) \psi_e  \nonumber \\
+\int d^3 x \psi_p^\dagger (\vec \alpha \cdot \frac 1i \vec
\partial +M_p\beta) \psi_p \nonumber \\
+\int d^3 x \frac 12[\vec E_\perp^2-\vec A_\perp \cdot
\partial_t^2\vec A_\perp +(\vec j_e-\vec j_p)\cdot
\frac 1{\vec \partial^2}\partial_t^2 \vec A_\perp] \nonumber \\
-\vec j_p \cdot \vec \partial \frac 1{\vec \partial ^2}(\vec
\partial \cdot \vec A) +j^0_e \partial_t \frac 1{\vec \partial ^2}
(\vec
\partial \cdot \vec A) \label{H'}
\end{eqnarray}
Here we have omitted self-energy terms which belong to the
issue of radiative corrections and the Lamb shift. In Eq.
(\ref{H'}), the first line is the usual electron
Hamiltonian $H_e$ in an electromagnetic field, the second
line is now a {\em free} proton Hamiltonian. It just gives
the proton rest mass if the proton were regarded as being
infinitely heavy compared to electron. The third line
(where $\vec E_\perp=-\partial_t\vec A_\perp$) is the
dynamic part of the electromagnetic field, which vanishes
for a stationary state (note that $\vec A_\perp=-\frac
1{\vec\partial ^2} \vec\partial \times\vec B$ is gauge
invariant, and is time-independent as $\vec B$ is). The
fourth line contains the gauge-dependent terms, which
vanish only in the Coulomb gauge. This is what we have
intended to show: in the stationary approximation, the
``electron energy'', $E_e$, computed in ({\em and only in})
the Coulomb gauge can label the total atomic energy (except
for the trivial proton mass term).

Beyond the stationary approximation, i.e., when considering
quantum fluctuations of the electromagnetic field, $E_e$ is
no longer a precise measure of atomic energy (as reflected
by the Lamb shift). Analogously, $\vec J_\gamma$ will be
non-zero in an atom, and $\vec J_p+\vec J_e$ receives
radiative corrections as well. Then, one may begin to
consider ``atomic spin structure''. In fact, Eq. (\ref{Ja})
is just the atomic version of the Jaffe-Manohar scheme of
separating the nucleon spin components. (The quark and
gluon angular momentum operators take the same forms as in
Eq. (\ref{Ja}), but with implicit color indices summed
over.) We see that from the perspective of constructing a
stationary angular-momentum eigenstate, the Jaffe-Manohar
scheme can indeed be useful, though limited to Coulomb
gauge.

The unsatisfactory aspect of the Jaffe-Manohar scheme, of course, is
that gauge dependence obscures the physical meaning of $\vec J_e$
and $\vec J_p$. The Ji scheme \cite{Ji97} is intended as an
improvement regarding gauge invariance. Its atomic version is
\begin{eqnarray}
\vec J_{\rm atom} &=& \int d^3 x \psi_e ^\dagger (\frac 12 \vec
\Sigma
+\vec x\times  \frac 1i \vec D_e)\psi_e \nonumber\\
&+& \int d^3 x \psi_p ^\dagger (\frac 12 \vec \Sigma +\vec x\times
\frac 1i \vec D_p)\psi_p \nonumber\\
 &+&\int d^3x \vec x \times (\vec E \times \vec B) \nonumber\\
 &\equiv& \vec J'_e+\vec J'_p+\vec J'_\gamma \label{Ja'}
\end{eqnarray}
The gauge invariance of $\vec J'_e$, $\vec J'_p$ and $\vec
J'_\gamma$ is evident by the use of the gauge-covariant
derivative in $\vec J'_e$, $\vec J'_p$ and the Poynting
vector in $\vec J'_\gamma$; therefore $\vec J'_e$, $\vec
J'_p$ and $\vec J'_\gamma$ have well-defined contents. But
relative to the art of spin decomposition as elaborated
above, (namely, to be able to choose a proper part which
can specify the quantum number of the whole atom,) the
gauge-invariance of $\vec J'_e$, $\vec J'_p$ and $\vec
J'_\gamma$ can also be a danger! The point is that without
any adjustable gauge variation to use, one can only {\rm
hope} that $\vec J'_e+\vec J'_p$ {\em intrinsically}
describe the atomic angular momentum. That hope, however,
cannot be realized. In fact, $\vec J'_e$ and $\vec J'_p$
are not angular-momentum operators at all: $\vec
J'_{e,p}\times \vec J'_{e,p}\neq i\vec J'_{e,p}$, thus
$\vec J'_e+\vec J'_p$ cannot possibly equal the proper
angular momentum operator, $\vec J_{\rm atom}$, except in
the trivial case of neglecting magnetic interaction. As a
cross-check, one can see that $\vec J'_\gamma$ is not an
angular-momentum operator either ($\vec J'_\gamma\times
\vec J'_\gamma\neq i\vec J'_\gamma$), and it does not
vanish even for a stationary system. $\vec J'_\gamma$
relates to $\vec J_\gamma$ by
\begin{equation}
\vec J'_\gamma =\vec J_\gamma +\int d^3x \vec x\times \rho \vec A.
\end{equation}
where $\rho$ is the total charge density. We have shown that, for a
stationary system, $\vec J_\gamma=0$ in the Coulomb gauge, while in
this gauge $\vec A=-\frac 1{\vec \partial^2} \vec j$ is not zero.
Therefore, despite being gauge-invariant, $\vec J'_e$ and $\vec J'_p$
are not useful in atomic physics (with regard to construction of
angular-momentum eigenstates), and $\vec J'_\gamma$ (more precisely,
$\vec x\times \rho \vec A$) represents a spurious angular momentum
of the electromagnetic field. If one were to regard $\vec J'_\gamma$ as
the electromagnetic angular momentum, then, even without considering
quantum fluctuations, the atomic spin would exhibit a non-trivial
structure: both the electron and photon would contribute to the atomic
spin and neither would be in a well-defined angular momentum eigenstate.
Such a structure, however, is just an artificial complication.

The recent proposal of Chen {\it et al} is a reconciliation of gauge-invariance
and construction of angular-momentum eigenstates. Its
atomic version is \cite{Chen08,Chen09,Chen11}:
\begin{eqnarray}
\vec J_{\rm atom} &=& \int d^3 x \psi_e ^\dagger [\frac 12 \vec
\Sigma
+\vec x\times  \frac 1i (\vec \partial -iq_e A_\parallel)]\psi_e \nonumber \\
&+& \int d^3 x \psi_p ^\dagger [\frac 12 \vec \Sigma +\vec x\times
\frac 1i (\vec \partial -iq_p \vec A_{\parallel})]\psi_p \nonumber\\
&+&\int d^3x [\vec E_\perp \times \vec A_\perp
+E_\perp^i\vec x\times \vec \partial A_\perp^i ] \nonumber \\
&\equiv& {\bf J}_e+{\bf J}_p+{\bf J}_\gamma   \label{BJa}
\end{eqnarray}
Here the longitudinal field $\vec A_\parallel=\vec \partial
\frac 1{\vec \partial ^2} (\vec\partial \cdot \vec A)$ is
the pure-gauge part of $\vec A$, and vanishes in the
Coulomb gauge. Its gauge transformation is the same as that
of the full $\vec A$, therefore $\vec \partial -iq \vec
A_\parallel$ is a (pure-gauge) covariant derivative.
Consequently, ${\bf J}_e$ and ${\bf J}_p$ are gauge
invariant, and in Coulomb gauge, ${\bf J}_e=\vec J_e$ and
${\bf J}_p=\vec J_p$. Analogously, ${\bf J}_\gamma$ is
gauge invariant, and is equal to $\vec J_\gamma$ in Coulomb
gauge. $\vec E_\perp=-\partial_t \vec A_\perp$ is the gauge
invariant, dynamical (transverse) part of the electric
field. It is evident from Eq. (\ref{BJa}) that ${\bf
J}_\gamma$ has the nice feature of vanishing identically
for a stationary configuration (while the gauge-dependent
$\vec J_\gamma$ does so in Coulomb gauge only). Regarding
the labeling of atomic states, ${\bf J}_e$ and ${\bf J}_p$
are therefore the pertinent and satisfactory operators to
use (in any gauge), just as ${\vec J}_e$ and ${\vec J}_p$
are in Coulomb gauge.

A gauge-invariant expression similar to Eq. (\ref{BJa}) can be
derived for the Hamiltonian, and indeed, for the whole
energy-momentum tensor, which can be put into the gauge-invariant,
canonical form:
\begin{equation}
T^{\mu\nu}=\bar \psi_e \gamma^\mu i\bar D_e^\nu\psi_e+\bar \psi_p
\gamma^\mu i\bar D_p^\nu\psi_p +F^{\rho\mu}\partial^\nu \hat A_\rho
-\eta^{\mu\nu} {\mathscr L} \label{T}
\end{equation}
Here $\bar D^\mu =\partial^\mu +iq\bar A^\mu$ is the
pure-gauge covariant derivative. $\bar A^\mu=-\partial^\mu
\frac 1{\vec
\partial ^2} (\vec\partial\cdot \vec A)$ is the pure-gauge part of
$A^\mu$. Its spatial component, $\vec {\bar A}$, is just $\vec
A_\parallel$. $\hat A_\rho=\frac 1{\vec
\partial^2}\partial_iF_{i\rho}=A_\rho -\bar A_\rho$ is the
(gauge-invariant) physical part of $A_\rho$, with spatial component
$\vec{\hat A}=\vec A_\perp$.

From Eq. (\ref{T}), the conserved four-momentum, $P^\nu
=\int d^3x T^{0\nu}$, is
\begin{equation}
P^\nu =\int d^3x(\psi_e^\dagger i\bar D_e^\nu \psi_e +\psi_p^\dagger
i\bar D_p^\nu \psi_p -E^i_\perp
\partial^\nu A^i_\perp -\eta^{0\nu} {\mathscr L})
\end{equation}
Here we have used the fact that $\int d^3x E^i_\parallel
\partial^\nu A^i_\perp=0$, where $\vec E_\parallel=-\vec \partial
\hat A^0$ is the gauge-invariant longitudinal part of the
electric field. In particular, the spatial three-momentum
is
\begin{equation}
\vec P =\int d^3x(\psi_e^\dagger \frac 1i\vec {\bar D}_e \psi_e
+\psi_p^\dagger \frac 1i\vec {\bar D}_p \psi_p +E^i_\perp \vec
\partial A^i_\perp) ,
\end{equation}
and the Hamiltonian is
\begin{eqnarray}
H_{\rm atom} &=&\int d^3x(\psi_e^\dagger i {\bar D}_e^0 \psi_e
+\psi_p^\dagger i\vec {\bar D}_p^0 \psi_p -E^i_\perp
\partial_t A^i_\perp -{\mathscr L}) \nonumber\\
&=&\int d^3 x \psi_e^\dagger (\vec \alpha \cdot \frac 1i \vec
D_e+M_e\beta +q_e\hat A^0) \psi_e \nonumber\\
&+&\int d^3 x \psi_p^\dagger (\vec \alpha \cdot \frac 1i \vec
D_p+M_p\beta +q_p\hat A^0) \psi_p \nonumber\\
&+&\int d^3 x \frac 12 (\vec E^2_\perp+\vec B^2-\vec E^2_\parallel)
\end{eqnarray}
The second line is the operator we used to replace $H_e$
for computing the atomic energy. \cite{Sun10} It is gauge
invariant, and equals $H_e$ in Coulomb gauge. Therefore,
the rest (also gauge-invariant) must intrinsically be
irrelevant for a stationary configuration. This property is
displayed by
\begin{eqnarray}
H_{\rm atom}=\int d^3 x \psi_e^\dagger (\vec \alpha \cdot \frac 1i
\vec
D_e+M_e\beta +q_e\hat A^0) \psi_e \nonumber\\
+\int d^3 x \psi_p^\dagger (\vec \alpha \cdot \frac 1i \vec {\bar
D}_p+M_p\beta) \psi_p\nonumber\\
+\int d^3 x \frac 12[\vec E_\perp^2-\vec A_\perp \cdot
\partial_t^2\vec A_\perp +(\vec j_e-\vec j_p)\cdot
\frac 1{\vec \partial^2}\partial_t^2 \vec A_\perp],
\end{eqnarray}
where the second line is a gauge-invariant free proton part, and the
third line vanishes for a static field. In deriving this expression,
we have rewritten $\vec B^2$ as
\begin{eqnarray}
\int d^3x \vec B^2&=&-\int d^3 xA^i_\perp \vec
\partial^2 A^i_\perp = \int d^3 xA^i_\perp (j^i_\perp-\partial_t^2
A_\perp^i)\nonumber\\
&=&\int d^3 x[j_\perp^i\frac{1}{\vec \partial ^2} (\partial_t^2
A^i_\perp-j_\perp^i)-A^i_\perp
\partial_t^2 A_\perp^i],
\end{eqnarray}
and self-energy terms have been discarded as above.

By examining the familiar and unambiguous examples in atomic
physics, we display clearly the art of spin decomposition: A good
decomposition should give a simple structure and physical picture,
and should not give arise to spurious complications.

We now turn to
the hadronic sector. A decomposition of QCD angular momentum
operator that respects the above art precisely mimics the atomic
expression in Eq. (\ref{BJa}) \cite{Chen11}:
\begin{eqnarray}
\vec J_{\rm QCD} &=& \int d^3 x \psi_q ^\dagger \frac 12 \vec \Sigma
\psi_q + \int d^3x \vec x \times \psi_q ^\dagger \frac 1i \vec {\bar
D}
\psi_q \nonumber \\
&+&\int d^3x ( -\mathcal D_t \vec {\hat A}) \times \vec {\hat
A}+\int d^3x \vec x \times
(-\mathcal D_t \hat A_i) \vec {\bar {\mathcal D}} {\hat A}_i \nonumber \\
&\equiv&{\bf S}_q +{\bf L}_q + {\bf S}_g +{\bf L}_g . \label{Chen}
\end{eqnarray}
These expressions are more complicated than Eq. (\ref{BJa})
due to color structure and non-linear terms. $\hat A^\mu$
is physical part of the non-Abelian gluon field. $\hat
A^\mu$ transforms in the same gauge-covariant manner as
$F^{\mu\nu}$, and therefore also requires covariant
derivatives. Note that in ${\bf S}_g$ and ${\bf L}_g$ the
pure-gauge field $\vec {\bar A}$ is used in $\vec {\bar
{\mathcal D}} \equiv \vec \partial -ig[\vec {\bar A}, ~]$,
while $\mathcal D_t\equiv \partial_t +ig[A^0,~]$ involves
the full $A^0$. As a result, ${\bf S}_g$ and ${\bf L}_g$
show a key difference from ${\bf S}_\gamma$ and ${\bf
L}_\gamma$, namely that they may survive in a stationary
configuration. This property may potentially be crucial at
the non-perturbative low-energy scale, and lead to a
sizable gluon contribution to the nucleon spin. In the
perturbative regime, however, the non-linear terms in ${\bf
S}_g$ and ${\bf L}_g$ are of higher order, and the
leading-order terms still vanish for a stationary
configuration. For example, if we consider a hadron made
entirely of heavy quarks (so that perturbative QCD
applies), at the order of one-gluon exchange, ${\bf
S}_g={\bf L}_g= 0$, and the hadron spin comes solely from
quarks. This picture is also true if the Coulomb gauge is
adopted for the (gauge-dependent) Jaffe-Manohar scheme
\cite{Jaff90}:
\begin{eqnarray}
\vec J_{\rm QCD} &=& \int d^3 x \psi_q ^\dagger \frac 12 \vec \Sigma
\psi_q + \int d^3x \vec x \times \psi_q ^\dagger \frac 1i \vec
\partial
\psi_q \nonumber \\
&+&\int d^3x \vec E \times \vec A+\int d^3x \vec x \times
E_i \vec \partial A_i \nonumber \\
&\equiv&{\vec S}_q +{\vec L}_q + {\vec S}_g +{\vec L}_g . \label{JM}
\end{eqnarray}
In a gauge other than Coulomb, however, ${\vec S}_g +{\vec L}_g$
develops a leading non-zero term of
\begin{eqnarray}
\int d^3x [(-\vec \partial A^0) \times \vec A+
 \vec x\times
(-\partial_i A^0) \vec \partial A_i] \nonumber\\
=\int d^3x A^0\vec x \times \vec
\partial (\partial_i A_i),
\end{eqnarray}
which can lead to another type of spurious gluon angular momentum in
a hadron. In this regard, it is somewhat awkward that many of the
theoretical techniques developed so far are for exploration of $\vec S_g$
in the light-cone gauge, which greatly simplifies the expression of
polarized gluon distribution function \cite{Jaff96}. In a very
recent paper \cite{Hatt11}, the formalism of gauge-field
decomposition in Ref. \cite{Chen08,Chen09,Chen11} is adopted to
construct a gauge-invariant gluon spin which agrees with $\vec S_g$
in light-cone gauge. While such an $\vec S_g$ may still reveal some
gluon dynamics in the nucleon, one must be very cautious in using
the data, since a sizable $\vec S_g$ so obtained does not
necessarily imply a significantly non-trivial gluonic content of the
nucleon. Analogously, if a sizable ${\vec J}'_g$ were found in the
Ji scheme \cite{Ji97}:
\begin{eqnarray}
\vec J_{\rm QCD} &=& \int d^3 x \psi_q ^\dagger \frac 12 \vec \Sigma
\psi_q + \int d^3x \vec x \times \psi_q ^\dagger \frac 1i \vec D
\psi_q \nonumber \\
&+&\int d^3x \vec x \times (\vec E \times \vec B)\nonumber \\
&\equiv&{\vec S}_q +{\vec L}'_q + {\vec J}'_g, \label{JM}
\end{eqnarray}
it may actually come from the
spurious gluon angular momentum $\int d^3 x \vec x\times
\psi^\dagger_q \vec A\psi_q$, and so may also not imply a significant
gluon content in the nucleon.

\acknowledgments

This work is supported by the National Science Foundation
of China under Grants No. 10875082 and No. 11035003, and by
the U.S. DOE under Contract No. DE-AC52-06NA25396. X.S.C.
is also supported by the NCET Program of the China
Education Department.

\end{document}